%% file: main.tex
\documentclass[conference]{IEEEtran}
\IEEEoverridecommandlockouts

\input{include/package.tex}
\input{include/command.tex}
\input{include/shorthand.tex}

\def\BibTeX{{\rm B\kern-.05em{\sc i\kern-.025em b}\kern-.08em
    T\kern-.1667em\lower.7ex\hbox{E}\kern-.125emX}}
\begin{document}

\title{ASTRA-sim2.0: Modeling Hierarchical Networks and Disaggregated Systems
for Large-model Training at Scale}
\input{include/author.tex}

\maketitle

\input{content/abstract.tex}


\input{content/introduction.tex}
\input{content/backgrounds.tex}
\input{content/motivation.tex}
\input{content/extensions.tex}
\input{content/case_studies.tex}
\input{content/related_work.tex}
\input{content/conclusion.tex}

\input{include/acknowledgement.tex}

\bibliographystyle{IEEEtran}
\bibliography{reference/reference.bib}

\end{document}

%% file: include/package.tex
\usepackage{cite}
\usepackage{amsmath,amssymb,amsfonts}
\usepackage{algorithmic}
\usepackage{graphicx}
\usepackage{textcomp}
\usepackage{xcolor}

\usepackage{lipsum}
\usepackage{xspace}
\usepackage{hyperref}
\hypersetup{colorlinks=true,linkcolor=blue,citecolor=blue,urlcolor=red}
\usepackage{comment}
\usepackage{caption}
\usepackage{subcaption}
\usepackage{tikz}
\usepackage{soul}
\usepackage{listings}

%% file: include/command.tex
\newcommand{\insertFigure}[5]{
    \begin{figure}[t]
      \centering
      \includegraphics[width=#3\linewidth]{figure/#1}
      \vspace{#4}
      \caption{#2}
      \label{fig:#1}
      \vspace{#5}
    \end{figure}
}

\newcommand{\insertFigureWide}[5]{
    \begin{figure*}[t]
      \centering
      \includegraphics[width=#3\linewidth]{figure/#1}
      \vspace{#4}
      \caption{#2}
      \label{fig:#1}
      \vspace{#5}
    \end{figure*}
}

\newcommand{\squishlist}{
 \begin{list}{$\bullet$}
  { \setlength{\itemsep}{0pt}
     \setlength{\parsep}{3pt}
     \setlength{\topsep}{3pt}
     \setlength{\partopsep}{0pt}
     \setlength{\leftmargin}{1.5em}
     \setlength{\labelwidth}{1em}
     \setlength{\labelsep}{0.5em} } }
     
\newcommand{\squishend}{
  \end{list}  }

\newcommand{\insertEqNoNum}[1]{%
\vskip -0.20in
\begin{gather*}
    #1
\end{gather*}
}

\newcommand*\circled[1]{\tikz[baseline=(char.base)]{\node[shape=circle,fill,inner sep=0.7pt] (char) {\textcolor{white}{#1}};}}

%% file: include/shorthand.tex
\newcommand{\astrasim}{ASTRA-sim\xspace}

\newcommand{\reducescatter}{Reduce-Scatter\xspace}
\newcommand{\allgather}{All-Gather\xspace}
\newcommand{\allreduce}{All-Reduce\xspace}
\newcommand{\alltoall}{All-to-All\xspace}

\newcommand{\ring}{Ring\xspace}
\newcommand{\fc}{FullyConnected\xspace}
\newcommand{\switch}{Switch\xspace}

\newcommand{\ringAlg}{Ring\xspace}
\newcommand{\directAlg}{Direct\xspace}
\newcommand{\hdAlg}{HalvingDoubling\xspace}

\newcommand{\tlarge}{Transformer-1T\xspace}
\newcommand{\gpt}{GPT-3\xspace}
\newcommand{\dlrm}{DLRM\xspace}


%% file: include/author.tex
\author{
\IEEEauthorblockN{
    William Won\IEEEauthorrefmark{1}\IEEEauthorrefmark{4}\thanks{\IEEEauthorrefmark{4}These authors contributed equally to this work.},
    Taekyung Heo\IEEEauthorrefmark{1}\IEEEauthorrefmark{4},
    Saeed Rashidi\IEEEauthorrefmark{1}\IEEEauthorrefmark{4},
    Srinivas Sridharan\IEEEauthorrefmark{2},
    Sudarshan Srinivasan\IEEEauthorrefmark{3},
    Tushar Krishna\IEEEauthorrefmark{1}
}
\IEEEauthorblockA{
    \IEEEauthorrefmark{1}Georgia Institute of Technology, Atlanta, GA, USA\\
    \IEEEauthorrefmark{2}Meta, Menlo Park, CA, USA\\
    \IEEEauthorrefmark{3}Intel, Bangalore, Karnataka, India
}
\IEEEauthorblockA{
    \IEEEauthorrefmark{1}\{william.won, taekyung, saeed.rashidi\}@gatech.edu, tushar@ece.gatech.edu\\
    \IEEEauthorrefmark{2}ssrinivas@fb.com \IEEEauthorrefmark{3}sudarshan.srinivasan@intel.com
}
\vspace{-2em}
}

%% file: content/abstract.tex
\begin{abstract}
As deep learning models and input data are scaling at an unprecedented rate, it is inevitable to move towards distributed training platforms to fit the model and increase training throughput. State-of-the-art approaches and techniques, such as wafer-scale nodes, multi-dimensional network topologies, disaggregated memory systems, and parallelization strategies, have been actively adopted by emerging distributed training systems. This results in a complex SW/HW co-design stack of distributed training, necessitating a modeling/simulation infrastructure for design-space exploration.
In this paper, we extend the open-source \astrasim infrastructure and endow it with the capabilities to model state-of-the-art and emerging distributed training models and platforms.
More specifically, (i) we enable \astrasim to support arbitrary model parallelization strategies via a graph-based training-loop implementation, (ii) we implement a parameterizable
multi-dimensional heterogeneous topology generation infrastructure with analytical performance estimates enabling simulating target systems at scale, and (iii) we enhance the memory system modeling to support accurate modeling of in-network collective communication and disaggregated memory systems.
With such capabilities, we run comprehensive case studies targeting emerging distributed models and platforms. This infrastructure lets system designers swiftly traverse the complex co-design stack and give meaningful insights when designing and deploying distributed training platforms at scale.
\end{abstract}

%% file: content/introduction.tex
\section{Introduction}\label{sec:Introduction}

The rapid growth in computation and memory requirement for Deep Neural Network (DNN) models is far greater than the performance and capacity scale of a single Neural Processing Unit (NPU, such as GPU or TPU). As an example, going from BERT~\cite{BERT} model to GPT-3~\cite{GPT3}, over the course of two years, requires 1800$\times$ more computation to train the model~\cite{CerebrasKeynote}. We have now reached the era of trillion parameter models~\cite{Transformer1T} that require 10's of terabytes of memory and zeta floating point  operations to train a model~\cite{CerebrasKeynote,Transformer1T}. Despite efforts to reduce the overhead of large model training on the workload side~\cite{shazeer2017outrageously}, big model training still remains challenging from the systems perspective~\cite{metafsdp}. Hence, distributed training is an inevitable option to keep up with the pace of increased resource requirements of DNN and Deep Learning (DL) training.

Designing an efficient distributed training system is challenging as there are many design choices such as parallelization strategies, NPU performance, NPU memory bandwidth, network topology, network bandwidth, and scheduling policies. Moreover, these design choices are interdependent, requiring the co-design of hardware and software for training platforms.
\astrasim~\cite{astrasim, AstraSimGithub} is an existing open-source infrastructure (originally developed by Georgia Tech, Intel and Meta). \astrasim aims to model the complete SW/HW co-design stack of distributed training systems, shown in \autoref{fig:AstraSimOverview}(a). It captures different aspects of distributed training platforms via three abstraction layers: (i) workload, (ii) system, and (iii) network. The workload layer implements the training loop (i.e., the DNN model, its parallelization strategy - data-parallel, model-parallel, etc., compute/communication ordering). The system layer provides various collective communication algorithm implementations (e.g., All-Reduce, All-to-All) and also manages pipelining and scheduling of communication operations. Finally, the networking layer models the HW/SW components of the network and simulates the traffic issued by the system layer.

\astrasim is a promising tool for exploring the design space of distributed training systems and has been leveraged by several recent works~\cite{themis,saeedACE, comet,hotiPaper,CCstudies,topology_alg_design}.
However, in this work, we identify limitations in \astrasim that restrict it from supporting arbitrary parallelism strategies, networks, and memory models. This comes from the rapidly changing SW/HW landscapes for DNN training as we describe next.

On the software end, there has been a growing interest in new parallelism strategies, both hand-designed such as 3D-parallelism~\cite{3DParallelMegatron,deepspeed}, FSDP~\cite{metafsdp}, ZeRO~\cite{rajbhandari2020zero}, expert parallelism~\cite{rajbhandari2022deepspeed} and discovered~\cite{flexflow, unity}. These strategies enable the training of large models, splitting datasets, parameters and optimizer state, while optimizing for communication~\cite{sapio2021scaling, li2019accelerating}. \astrasim did not have a strong motivation to support arbitrary parallelism when it was proposed as there were a handful of parallelism strategies such as data parallelism~\cite{PytorchDistributedDP}, model parallelism, and hybrid~\cite{shoeybi2019megatron}.

\insertFigureWide{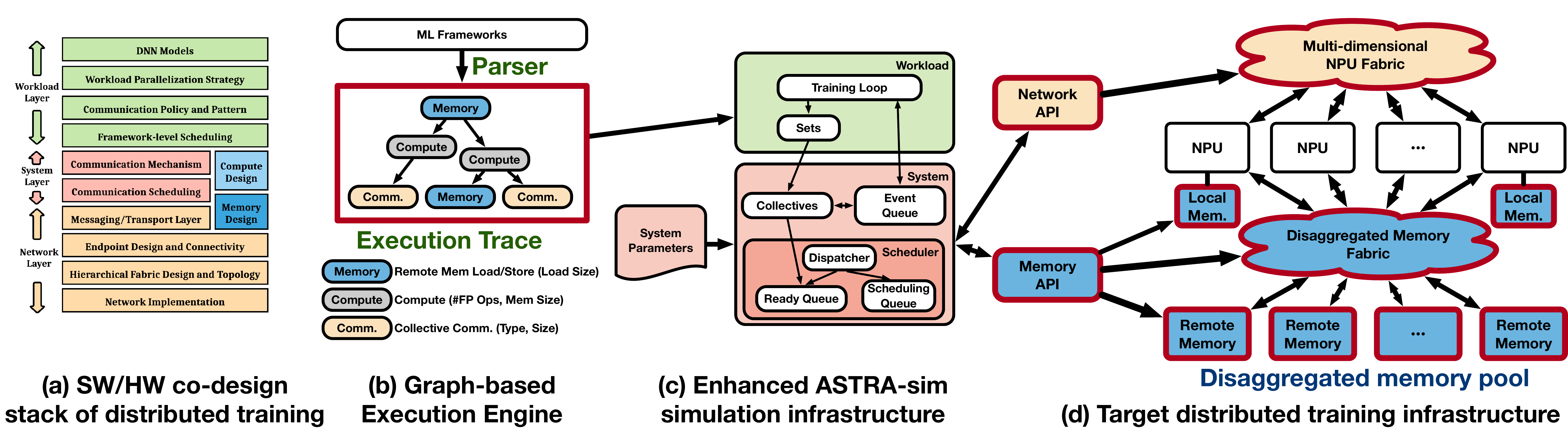}{
Overview of Proposed Infrastructure for Modeling Next-generation Training platforms.
The components extended in \astrasim2.0 from the original \astrasim to model emerging platforms are marked in bold.
}{0.98}{-2mm}{-5mm}

The hardware landscape for distributed training has been evolving rapidly as well. State-of-the-art systems extensively deploy multi-dimensional network topologies with hierarchical bandwidths to interconnect NPUs~\cite{nvidiadgx, dgxa100, tpuarch, cloudtpuarch, intelpontevecchio}. This is because increasing the aggregated network BW per NPU through a single dimension is fundamentally limited by the link technology the network is leveraging (e.g., current NVLink~\cite{nvlinkBridge} offers up to 450 GB/s). Naively scaling out through NIC is also not practical due to engineering limitations such as dollar-cost, power, and thermal problems. Meanwhile, wafer-scale systems~\cite{cerebraswhitepaper, dojo} tackle the communication problem by fabricating NPU chiplets on a large-wafer with low-dimensional, high on-chip networking, then scaling out such wafers using NICs. In order to study these technology-driven network landscapes, there is a need for a mechanism to represent and study arbitrary multi-dimensional topologies at scale, with different shapes and BW configurations.
\astrasim natively uses the Garnet simulator~\cite{niket2009garnet} from gem5 as its network layer, which has limitations in modeling such platforms.

Memory disaggregation, which allows GPUs to access a larger remote memory pool, is another
promising HW solution to overcome the limited GPU memory capacity per node. Although the concept has been studied for several decades~\cite{comer1990new, iftode1993memory, lim2009disaggregated, gu2017efficient, aguilera2017remote, shan2018legoos, ruan2020aifm, guo2022clio}, the network and memory did not support memory disaggregation. Motivated by the need for memory disaggregation, the computing industry is now building a framework with a new network technology, compute express link (CXL)~\cite{cxl}.
As distributed training systems will benefit from disaggregated memory, there is a strong need for exploring this design space.
\astrasim uses a simple BW number to model memory and cannot capture this complex design-space.

In this work, we address the aforementioned limitations of \astrasim and enhance it via three novel features, as shown in \autoref{fig:AstraSimOverview}(b)-(d): (i) arbitrary parallelism support, (ii) hierarchical network support, and (iii) memory model support.
We add arbitrary parallelism support by encoding parallelism strategies as execution traces and developing a parser to translate these into compute and communication tasks with dependencies.
For network support, we developed a taxonomy to define hierarchical topologies and created an analytical model to estimate performance when running a topology-aware collective over the physical topology. For the memory models,
we augment \astrasim
with the ability to model local (e.g., HBM) and networked remote (pooled) memories.

Using these enhancements, we present case studies to deliver key insights about future platforms.
We compared conventional multi-dimensional and wafer-scale systems and found that with appropriate collective scheduling and parallelization strategy designs, conventional systems can match wafer-scale systems' performance, whereas wafer-scale shows up to 2.51$\times$ better collective time when scaled.
We also compared disaggregated memory architectures and found that communication time dominates in training a Mixture-of-Experts (MoE) model, and identify configurations that can hide communication time to provide 4.6x speedup over a baseline Zero-Infinity~\cite{rajbhandari2021zero}.

%% file: content/backgrounds.tex
\section{Background}

\subsection{Distributed Training}\label{subsec:DistributedTraining}
\circled{1} \textbf{Synchronous/Asynchronous Training.}
When models/data are distributed across NPUs, it is crucial to decide when and how to synchronize such distributed information across them. The asynchronous training approach, as the name suggests, communicates among NPUs in an asynchronous manner. Therefore, asynchronous training suffers from the convergence problem~\cite{saeedACE} and is more complex to implement and maintain~\cite{distributedSurvey}. Therefore, the most common approach is synchronous distributed training. In this mechanism, all nodes work on their own data and synchronize the distributed information altogether before proceeding to the next iteration, usually in the form of collective communications~\cite{astrasim, li2019accelerating}.

\circled{2} \textbf{Parallelization Strategy.}
Each parameter, including model weights and input data, is distributed across NPUs. How each parameter is sharded and distributed is dictated by its ruling parallelization strategy~\cite{astrasim}. The three most pervasive parallelism strategies are: data-parallel (DP), model-parallel (MP), and pipeline-parallel (PP). DP distributes mini-batch across NPUs and synchronizes weight gradients during the backward pass~\cite{astrasim, distributedSurvey}. MP, on the other hand, distributes a model evenly across NPUs and communicates forward activation and input gradients pass~\cite{distributedSurvey}. MP and DP are orthogonal patterns, therefore MP and DP can be used simultaneously, called hybrid-parallel scheme~\cite{shoeybi2019megatron}. PP distributes model layers across nodes and processes micro-batches in a pipelined manner~\cite{huang2019gpipe, harlap2018pipedream}. Other parallelization strategies are also actively being investigated~\cite{rajbhandari2020zero, rajbhandari2021zero, ren2021zero}.

\circled{3} \textbf{Training Loop.}
In addition to parallelization, the order of communication and computation must be clearly defined to execute distributed training. Such computation and communication ordering information is named a training loop~\cite{astrasim}.

\subsection{Collective Communication}\label{subsec:CollectiveCommunication}

\circled{1} \textbf{Collective Communication.}
Depending on the parallelization strategy, models and/or input batches are distributed across NPUs. Therefore, it is unavoidable that devices should communicate and synchronize data, such as forward activation or weight/input gradients~\cite{themis}. This traffic is commonly formulated and processed in the form of collective communications. Some common collective patterns in distributed training are shown in \autoref{fig:CollectiveDefinition}. With a synchronous training approach, the most pervasive collective pattern is \allreduce~\cite{dlCollective}, which could be logically viewed as \reducescatter\ followed by an \allgather.

\circled{2} \textbf{Hierarchical Collective Algorithm.}
There exist several basic topology-aware collective communication algorithms to execute these communication patterns. A handful of examples of basic topology-aware \allreduce collective algorithms include Ring-based~\cite{ringCollective}, Tree-based~\cite{treeAlg}, and Halving-Doubling~\cite{rabenseifner}. However, when the underlying network topology is multi-dimensional, such basic algorithms would not perform optimally as the logical topology each algorithm assumes mismatches the physical one. In order to mitigate such an effect, multi-rail hierarchical collective algorithms have been proposed~\cite{blueconnect}. Using this scheme, in order to run an \allreduce collective on an $N$-dimensional topology:
\squishlist
    \item Run \reducescatter in ascending order from Dim 1, then Dim 2, $\cdots$, up to Dim $N$.
    \item Run \allgather in descending order from Dim $N$, $\cdots$, Dim 2, down to Dim 1.
\squishend

\insertFigure{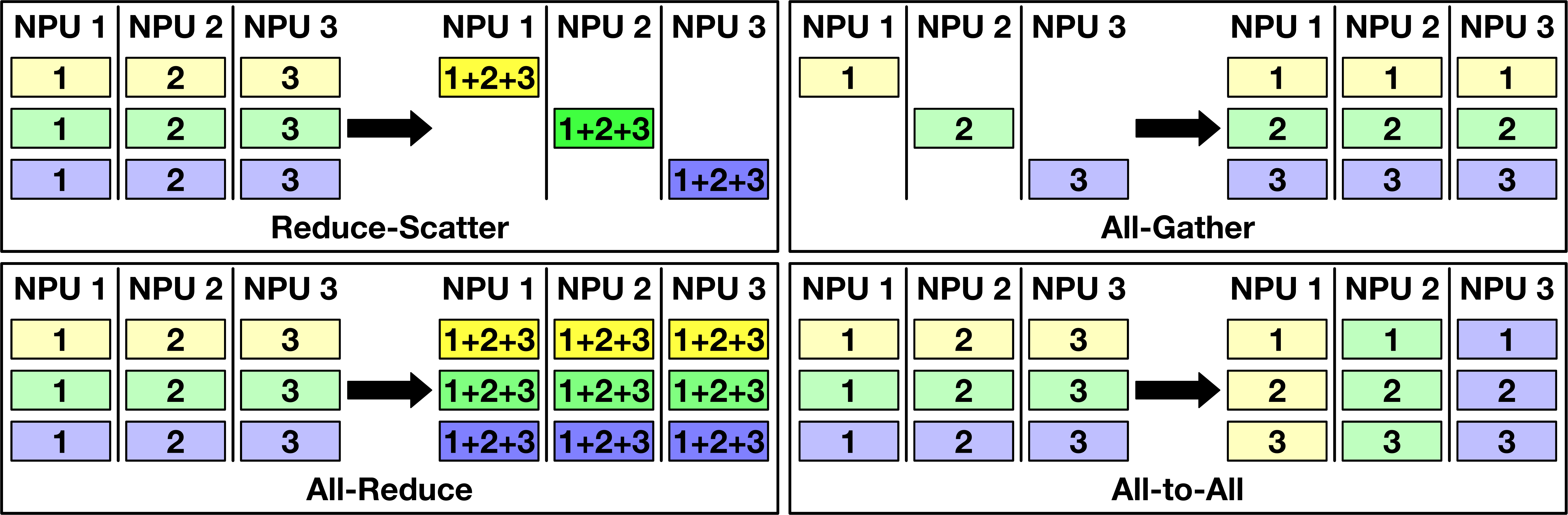}{
Definition of \reducescatter, \allgather, \allreduce, and \alltoall collective communication patterns.
}{1}{-5mm}{-4mm}

\subsection{ASTRA-sim}
Vast design choices of distributed training shown in \autoref{subsec:DistributedTraining}, combined with diverse hardware configurations create an enormous SW/HW design space of distributed training as depicted in \autoref{fig:AstraSimOverview}(a). Such enormous design space cannot be solely explored by only leveraging physical systems, especially at scale. Therefore, a simulation-based mechanism to quickly model and profile distributed training platforms is necessary for design-space exploration.
\astrasim~\cite{astrasim} is a distributed training simulation framework to exactly address this demand. \astrasim captures the training configuration explained in \autoref{subsec:DistributedTraining}
Its high-level components are summarized in \autoref{fig:AstraSimOverview}(c).
It codifies the complex SW/HW search space across three abstraction layers. The workload layer lets the user describe and define target DNN models, target parallelization strategies, and training loops.
The system layer implements collective communication algorithms, schedules compute and communication operations, and manages compute-communication overlap. Compute times are fed in via external NPU models~\cite{samajdar2018scale} or real system measurements. Communication times are computed using a network simulator. The default simulator is Garnet~\cite{niket2009garnet} from gem5.
It reports detailed system and network-level behaviors as well as end-to-end training throughput.

%% file: content/motivation.tex
\section{Motivation}
Even though \astrasim framework has allowed brisk navigation of distributed training search space~\cite{themis, saeedACE}, the tool as-is does not meet the demand to capture more complex target platforms. In this section, we motivate the need to extend the \astrasim toolchain to enable modeling state-of-the-art and futuristic training systems.
Specifically, we identify three emerging requirements for \astrasim as shown below.
\begin{itemize}
    \item Ability to model arbitrary parallelisms
    \item Ability to model multi-dimensional hierarchical networks
    \item Ability to model memory systems
\end{itemize}

\subsection{Ability to Model Arbitrary Parallelism Strategies}
One of the major limitations of \astrasim is the limited parallelism support. The original \astrasim cannot support complex parallelization strategies such as pipeline parallelism~\cite{harlap2018pipedream, huang2019gpipe} and 3D parallelism~\cite{deepspeed}. There are two reasons for the limitation. First, \astrasim assumes that all NPUs perform the same operation at the same time. While this assumption saves the engineering overhead in implementing data parallelism and model parallelism, it does not allow pipeline parallelism as it requires executing different operations on each NPU at the same time. Second, parallelization strategies are tightly coupled with the frontend implementation of \astrasim and implemented as separate training loops in the workload layer. Parallelization strategies for distributed training are an active area of research~\cite{metafsdp, dean2012large, sergeev2018horovod, harlap2018pipedream, huang2019gpipe, lepikhin2020gshard}, and sometimes several strategies are jointly applied~\cite{rajbhandari2022deepspeed}. Therefore, to evaluate arbitrary parallelism strategies, it is critical to decouple parallelization strategies from the \astrasim implementation.

\subsection{Ability to Model Multi-dimensional Networks}
From the necessity to distribute and synchronize models and data across devices, large-scale distributed training is usually communication-bound~\cite{sapio2021scaling, dlCollective, li2019accelerating}. Therefore, in order to maximize training performance, state-of-the-art systems mix and match a plethora of networking technologies. This usually ends up in a system having multi-dimensional network topologies with heterogeneous bandwidth configurations~\cite{themis}. As an instance, NVIDIA DGX-A100~\cite{dgxa100} exploits a 2-dimensional network topology whose first dimension is NVIDIA NVLink~\cite{nvlinkBridge} then scaled-out using InfiniBand~\cite{MellanoxSHARP} or Ethernet~\cite{NIC400G, NIC800G} technologies. The Google Cloud TPUv4~\cite{tpuarch} leverages a 3D Torus where each inter-core interconnect runs at 448 Gb/s~\cite{tpuv4ici}.

Although \astrasim can, in principle, target multi-dimensional networks, it only supports a limited set of pre-defined network topologies -- 2D and 3D torus.
In order to study different topologies, one must implement both a new network topology in Garnet and its corresponding topology-aware hierarchical collective algorithm, which significantly drags \astrasim's strength of swift distributed training system modeling and performance analysis.

Therefore, it is necessitated to attach a more powerful network backend to the \astrasim framework for rapid design-space exploration of state-of-the-art and futuristic training platforms. It must define a systematic mechanism to represent arbitrary multi-dimensional network topologies at scale. With such notation, the user cam swiftly represent an arbitrary multi-dimensional networks, instead of manually implementing network topology files and their corresponding collective communication algorithms.

\subsection{Ability to Model Emerging Memory Systems}
As DNN model parameters have to be loaded from and stored back to memory, having an efficient memory system is critical in distributed training. To design an efficient memory system, exploring the memory system design space is essential. However, as the original \astrasim does not have detailed memory models, it limits the opportunity to explore the design space. We find that \astrasim should support the following three features. The first feature is the ability to model local HBM memory. \astrasim should have a local memory model that allows how the performance changes as HBM latency and bandwidth vary. This feature allows system and architecture designers to find the optimal local HBM configuration within the same budget. The second feature is the support for memory disaggregation. It is well known that the limited capacity of GPUs is the major bottleneck in large-model training. Model parallelism~\cite{distributedSurvey} and memory optimizations~\cite{rajbhandari2020zero, ren2021zero, rajbhandari2021zero} have been widely adopted to overcome the limitation. While the proposed solutions have been effective in reducing per-GPU memory footprint, they come with critical limitations such as increased computation and communication time. Memory disaggregation is a fundamental solution to overcome the NPU memory capacity limitation by allowing NPUs to access a larger remote memory pool. Emerging interconnects such as CXL~\cite{cxl} accelerate this trend. \astrasim should be able to answer research questions such as the optimal configurations and design for memory disaggregation. The last feature is in-switch collective communication support. With the introduction of memory disaggregation, network switches are introduced in the memory access path of training systems. Performing collective communication in switches is an attractive option to improve the performance of distributed training by reducing communication time~\cite{li2019accelerating, gebara2021network, sapio2021scaling, lao2021atp, de2021flare, pan2022libra}. To find out the performance benefit and trade-offs of in-switch collective communication, \astrasim should support in-switch collective communication.

%% file: content/extensions.tex
\begin{figure}[!t]
\begin{minipage}{\linewidth}
\begin{lstlisting}
eg = None
if args.eg:
    eg_file = f"{out_file_prefix}_eg.json"
    eg = ExecutionGraphObserver()
    eg.register_callback(eg_file)
    eg.start()
...
if eg:
    eg.stop()
    eg.unregister_callback()
\end{lstlisting}
\captionof{lstlisting}{Execution trace collection example~\cite{metaparam}.}
\label{snippet:param-example}
\end{minipage}
\end{figure}

\section{Extensions to \astrasim}
In this section, we introduce the new features we added to \astrasim and describe how they are implemented.
All extensions are released and publicly accessible in the \astrasim repository\footnote{https://github.com/astra-sim/astra-sim}.

\subsection{Graph-based Execution Engine}
\label{sec:eg}
To support arbitrary parallelization strategies, we replace the frontend of \astrasim with a graph-based execution engine. The graph-based execution engine decouples parallelization strategies from the frontend implementation. As the name implies, the graph-based execution engine works on input graphs. The input graphs encode the execution of ML models and their associated parallelization strategies, which can be generated from ML frameworks such as PyTorch~\cite{executiongraph}, TensorFlow~\cite{tensorflow2015-whitepaper}, and FlexFlow~\cite{flexflow}. Code snippet \ref{snippet:param-example} presents how graphs can be collected with PyTorch. PyTorch offers a seamless option for collecting such graphs, which does not require any modifications to the model. The collected graphs are named execution traces (ETs). ETs are fed into the frontend, and the execution engine is responsible for simulating a distributed training system. ETs encode critical information for simulation such as memory access, computation, and communication. Each operation is modeled as a node, and their dependencies are presented as edges as shown in \autoref{fig:AstraSimOverview}(b). In ETs, parallelization strategies are encoded with dependencies. As each NPU has an independent graph-based execution engine, each NPU can run different operations. The engine consumes nodes one by one, and the dependent nodes become ready to be issued when all of their parent nodes are completed. Nodes are completed after a specific delay, and the delay is determined by the node type and metadata. The execution engine continues the simulation until it consumes all nodes.

We define a common format for execution traces, called \astrasim ET, to avoid implementing ET parsers for all different ET types in \astrasim. Instead, we provide a converter from any ET (e.g., PyTorch ET) to \astrasim ET\footnote{Currently, PyTorch and FlexFlow are supported.}. \astrasim ET has three node types: compute, memory, and communication as presented in \autoref{fig:AstraSimOverview}(b). Each node has metadata that is critical for simulating the operation. Compute nodes have the tensor size and the number of floating point operations to perform computation. \astrasim calculates the number of cycles to perform the operation with an internal roofline model. Memory nodes measure the number of cycles to store or load a tensor. Therefore, the nodes have a tensor size as metadata. Communication nodes encode the communication type (collective communication between NPUs or peer-to-peer communication between a pair of NPUs) and the communication size. This information gets translated into a network delay by the underlying system and network layers of \astrasim.

\subsection{Multi-dimensional Network Representation}\label{subsec:MultiDimNetwork}

\insertFigure{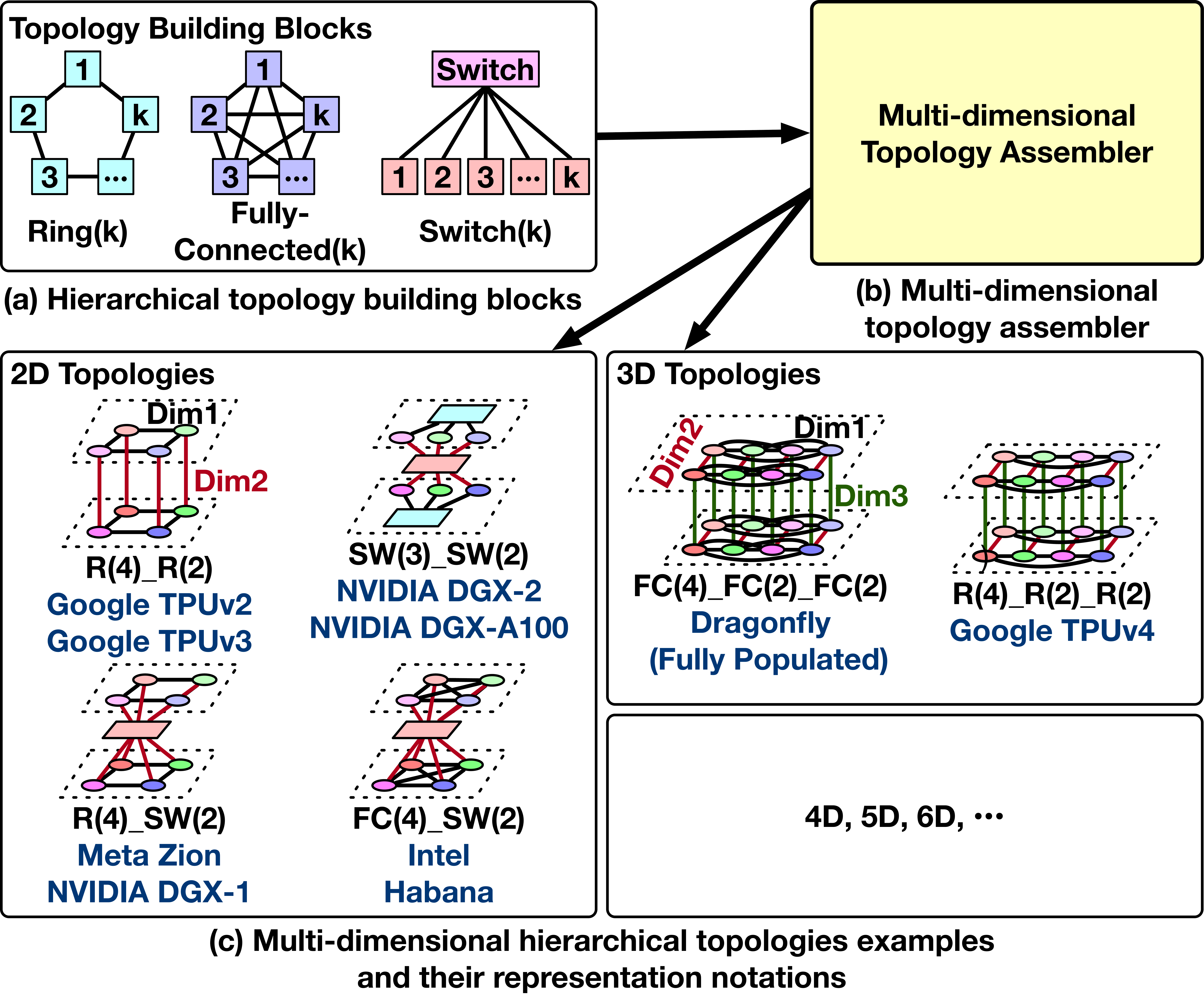}{
(a) Hierarchical topology building blocks: \ring, \fc, and \switch (b) Multi-dimensional network topologies are created by stacking up network building blocks (c) Multi-dimensional hierarchical topology examples, their shape notations, and corresponding distributed training framework.
}{1}{-1.5em}{-0.5em}

In order for users to quickly target arbitrary multi-dimensional network topologies, it is crucial to design a generic notation to represent such multi-dimensional shapes. In this paper, we propose a taxonomy that constructs a multi-dimensional topology by stacking up network building blocks in a hierarchical manner. \autoref{fig:MultiDimNetwork}(a) shows the three network building blocks utilized in this paper: \ring (R), \fc (FC), and \switch (SW). Ring($k$) connects $k$ NPUs in a ring shape (i.e., two connections per every NPU). FullyConnected($k$), on the other hand, offers all-to-all connectivity among all pairs of NPUs. Finally, Switch($k$) connects all $k$ NPUs using an external switch fabric. We chose these three as the network building blocks as they have corresponding well-known topology-aware collective algorithms as summarized in \autoref{table:TopologyAwareCollective}\footnote{Even if the underlying system uses other topologies, they are logically reduced into one of these building blocks due to the collective communication library~\cite{nccl, oneCCLDoc}. This is a unique feature of DL training platforms.}.

\input{table/TopologyAwareCollective.tex}

Multi-dimensional topologies can be generated by assembling these blocks in an arbitrary hierarchical manner, as glimpsed in \autoref{fig:MultiDimNetwork}(b). A handful of example constructed topologies are shown in \autoref{fig:MultiDimNetwork}(c). Ring(4)\_Ring(2) simply denotes a 2D Torus with 8 NPUs in total, where the first dimension is Ring(4) and two such Dim 1 networks are interconnected using Ring(2) topology. Ring(4)\_Switch(2), on the other hand, has the same Dim 1 but planes are being scaled out using an external switch instead. An example 3D topology from \autoref{fig:MultiDimNetwork}(c) is FC(4)\_FC(2)\_FC(2), a fully-populated DragonFly~\cite{dragonfly} topology with 16 NPUs. Ring(4)\_Ring(2)\_Ring(2) is also shown, where the NPU placement is equivalent but topologies connecting them are substituted to \ring, thereby resulting in a 3D torus instead. The number of network dimensions or the building blocks' order is not restricted, thus arbitrary 4D, 5D, $\cdots$, networks can easily be represented using the same notation. Note that each and every example topologies listed in \autoref{fig:MultiDimNetwork}(c) corresponds to some state-of-the-art distributed training platforms, demonstrating the power of our proposed representation in modeling the design space.

With this representation, designing a multi-dimensional topology-aware collective is straightforward and requires minimal modification. As explained in \autoref{subsec:CollectiveCommunication}, multi-rail hierarchical collective algorithms can be run by iteratively running the basic topology-aware collective algorithm on each dimension. Recall that we deliberately chose network building blocks that have known congestion-free collective algorithms. The corresponding topology-aware collective algorithms are listed in \autoref{table:TopologyAwareCollective}. Consequently, collective communications on any arbitrary multi-dimensional network can be run by running these basic algorithms in order and requires no further modification.

\subsection{Analytical Network Backend}

Supporting arbitrary multi-dimensional network topologies shown in \autoref{subsec:MultiDimNetwork}, we implemented a new analytical network backend\footnote{https://github.com/astra-sim/analytical} and ported it to the \astrasim framework. The following points summarize 
why an analytical equation-based network was sufficient for our purpose:
\squishlist
    \item There is a need for first-order design-space exploration (topology shape and BW) of the target system at scale.
    \item As Garnet is most suitable for modeling network-on-chip targets, it is challenging to easily model arbitrary multi-dimensional network topologies, as discussed in \autoref{subsec:MultiDimNetwork}.
    \item Given the scale (1000s of NPUs) of state-of-the-art and futuristic systems and DL models, cycle-level simulation using Garnet is too slow to be practical.
    \item Multi-dimensional topologies run a topology-aware multi-rail hierarchical collective algorithm, which does not create any network congestion. Thanks to this effect, analytical equation-based modeling shows marginal accuracy change over cycle-accurate simulations, and in fact closely matches real system measurements for a small system, as we show later.
\squishend

\begin{figure}[!t]
\begin{minipage}{\linewidth}
\begin{lstlisting}
sim_schedule(delta, callback)
sim_send(msg_size, dest, callback)
sim_recv(msg_size, src, callback)
\end{lstlisting}
\captionof{lstlisting}{Abstract view of example \astrasim frontend NetworkAPI methods~\cite{hotiPaper}.}
\label{snippet:networkAPI}
\end{minipage}
\end{figure}

In order to model a communication between two NPUs, ASTRA-sim frontend delegates the network backend to simulate such a communication and requests the backend to invoke a callback function to notify the transmission is completed. This protocol is defined in the form of NetworkAPI (\autoref{fig:AstraSimOverview}(c)) methods~\cite{hotiPaper}. Several examples of NetworkAPI methods are shown in Snippet \ref{snippet:networkAPI}. Whenever a communication request, such as \texttt{sim\_send} or \texttt{sim\_recv} is initiated, the analytical network backend leverages a simple equation to estimate the communication delay instead of simulating actual network behaviors:

\insertEqNoNum{
    \text{Time} = (\text{LinkLatency} \times \text{Hops}) + \frac{\text{MessageSize}}{\text{LinkBandwidth}} \label{eq:AnalyticalEq}
}
and simply schedules the callback function to be invoked after this delay, unlike the original Garnet backend which runs packet-level cycle-accurate simulations\footnote{
This approach may have limitations when the network contains non-trivial behaviors, such as network congestion or link oversubscription.
Implementing first-order congestion modeling into the analytical backend is our future work.
}. Modeling communication with serialization and link delay is suitable when the communication size is relatively large to be bandwidth-bound (e.g., DLRM and Transformer-1T has 100MB--1GB collectives). The analytical equation could be amended to consider other effects, such as wire propagation delay, as desired. For example, complex system and network optimizations (such as remote memory management or in-switch collective communication) can be captured by equations (\autoref{sec:memory-models}). 

\insertFigure{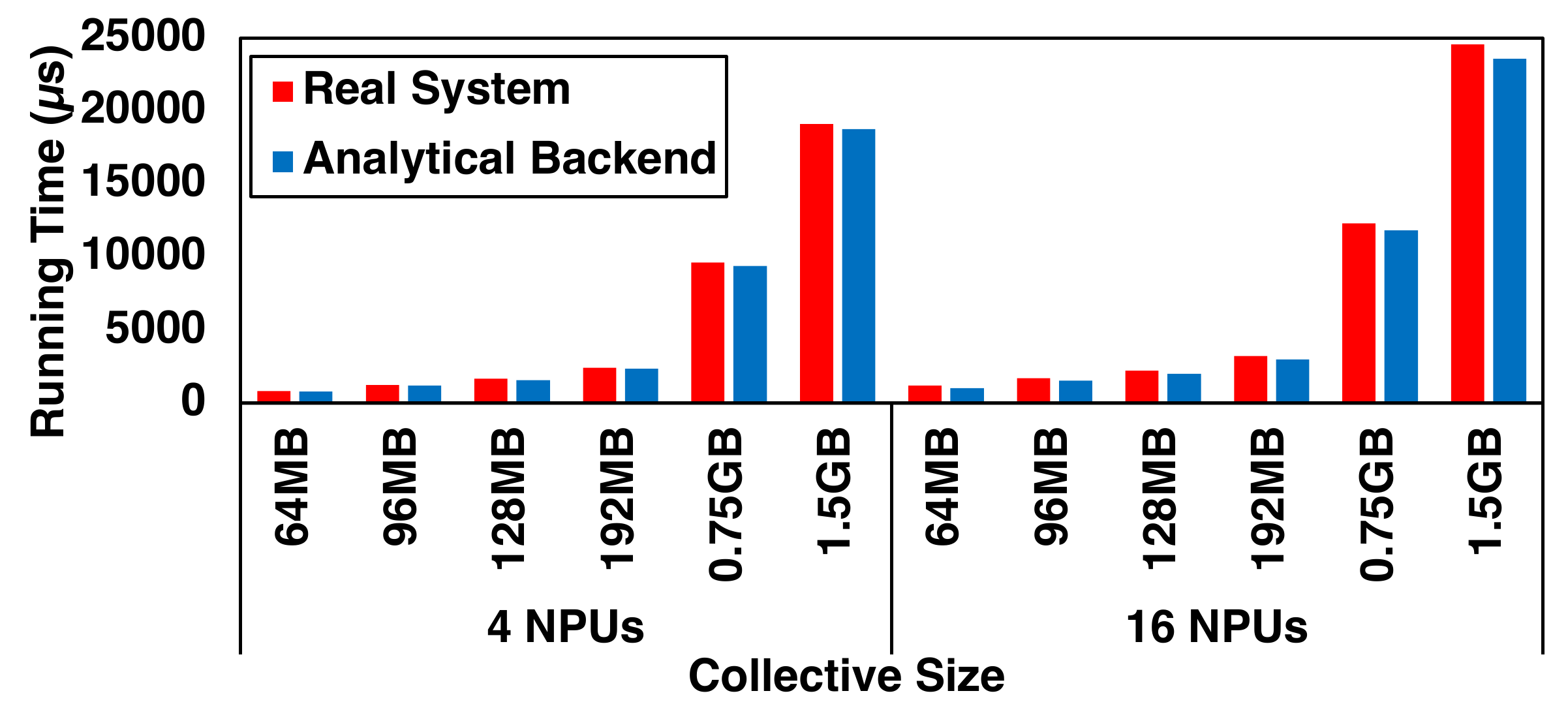}{
Analytical network backend validation over real system measurements ranging from 64MB--1.5GB \allreduce collectives.
}{0.95}{-3mm}{-5mm}

\textbf{Validation.}
In order to show the accuracy of the analytical network backend, we constructed two real systems and compared various-sized \allreduce running time. The two real systems leverage NCCL v2.4.6~\cite{nccl} which consist of 4 and 16 NVIDIA V100 GPUs~\cite{nvidiav1002017} using a \ring topology with 150 GB/s NVLink~\cite{nvlinkBridge} among GPUs. \autoref{fig:RealSystemValidation} shows the result. We run 64 MB -- 1.5 GB \allreduce and the results suggest the mean error of simulation over all configurations is 5\%.

\textbf{Speedup.}
In order to measure the simulation time improvement, we run a 1MB \allreduce simulation on a 3D Torus with 64 NPUs (4$\times$4$\times$4). On Garnet-based \astrasim, the simulation took 21.42 minutes to finish. For the same configuration, the analytical backend only spent 1.70 seconds, showing 756$\times$ speedup in simulation runtime. Further, the analytical backend supports a 3D Torus with 4K NPUs (16$\times$16$\times$16) in just 3.14 seconds. Nearly three-orders-of-magnitude speedup proves the capabilities of analytical network backend to profile systems of scale at speed.

\subsection{Memory Models}
\label{sec:memory-models}
We add a memory API to \astrasim to support various memory models as shown in \autoref{fig:AstraSimOverview}(d). The goal of memory API is to model various memory systems ranging from local memory to disaggregated memory. Memory API takes tensor location (local or remote), tensor size, memory bandwidth, and memory system design as arguments and returns the number of cycles to load or store a tensor to a memory system. Tensor size and location are encoded in the metadata of ET nodes, and memory bandwidth and system design are given as system configurations. Memory API supports local memory, remote memory, and in-switch collective communication. Memory API determines the model to run based on the tensor location and system parameters.

\begin{figure}[t!]
    \centering
    \begin{subfigure}{0.4\linewidth}
        \center
        \includegraphics[width=\linewidth]{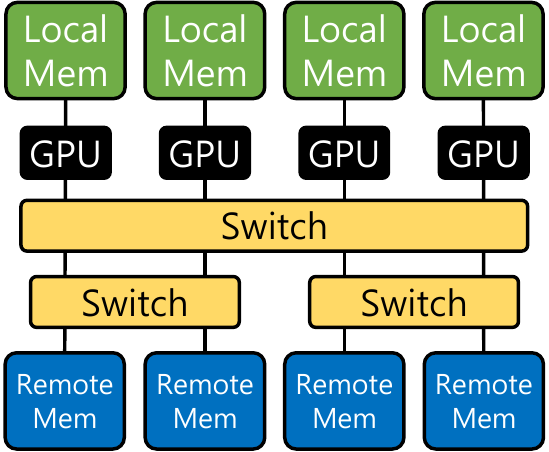}
        \\
        \caption{Multi-level switches.}
        \vspace{0.5em}
    \end{subfigure}
    \hspace{0.5em}
    \begin{subfigure}{0.4\linewidth}
        \center
        \includegraphics[width=\linewidth]{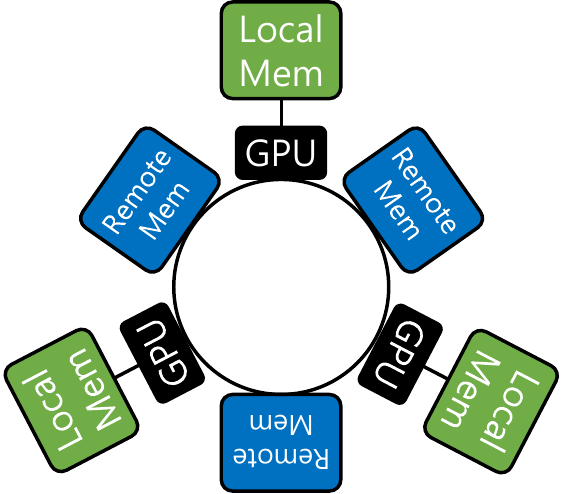}
        \\
        \caption{Ring.}
        \vspace{0.5em}
    \end{subfigure}
    \\
    \begin{subfigure}{0.4\linewidth}
        \center
        \includegraphics[width=\linewidth]{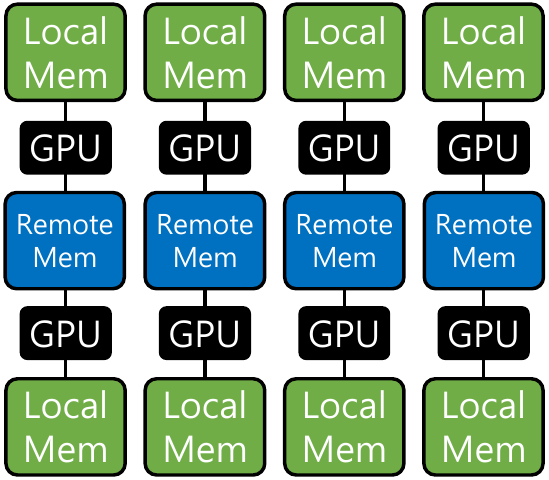}
        \\
        \caption{Mesh.}
    \end{subfigure}
    \hspace{0.5em}
    \begin{subfigure}{0.4\linewidth}
        \center
        \includegraphics[width=\linewidth]{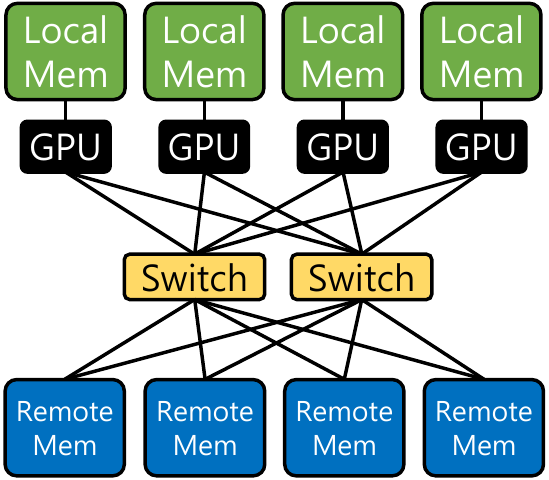}
        \\
        \caption{Hierarchical.}
    \end{subfigure}
    \\
    \caption{Various memory pool architectures.}
    \label{fig:memory-pool-architectures}
    \vspace{-2em}
\end{figure}

\circled{1} \textbf{Local Memory Model.} This is a simple memory bandwidth model with memory access latency, tensor size, and memory bandwidth as presented in an equation below. Memory access latency and memory bandwidth are given to \astrasim as system parameters, and payload size is encoded in a memory access node of an ET.

\begin{gather*}
\label{eq:remote-mem-access-time}
\scalebox{.9}{$
    \begin{aligned}[b]
    &(Memory\ Access\ Time) \\
    &= (Memory\ \ Access\ Latency) \\
    &+ (Tensor\ Size) / (Memory\ Bandwidth)
    \end{aligned}$}
\end{gather*}

\circled{2} \textbf{Remote Memory Model.} This model has the ability to calculate the data transfer time with a disaggregated memory system. In addition to the default parameters for the local memory model, this model takes the disaggregated memory design as a parameter. A disaggregated memory can take any design such as multi-level switches, rings, mesh, and hierarchical as shown in \autoref{fig:memory-pool-architectures}. Different design choices result in different data transfer times because the load on links and the number of network hops change.

\begin{figure}[t]
    \centering
    \includegraphics[width=\linewidth]{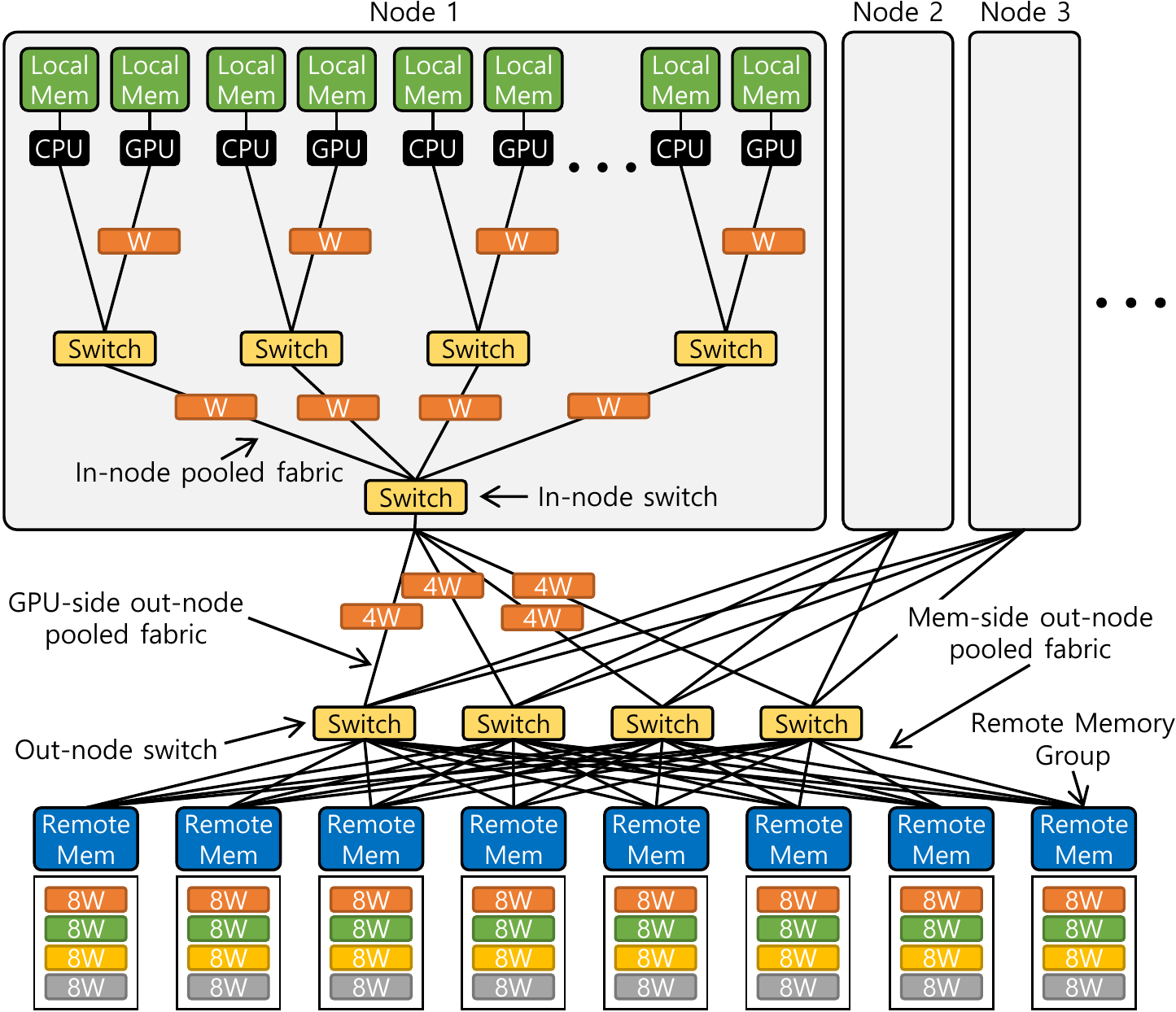}
    \caption{Remote memory model illustration.}
    \label{fig:remote-memory-model-illustration}
    \vspace{-1em}
\end{figure}

The remote memory model calculates the data transfer time for a given disaggregated memory system for given parameters. For ease of explanation, let's assume a system with a hierarchical disaggregated memory. \autoref{fig:remote-memory-model-illustration} illustrates how the remote memory model works for the hierarchical disaggregated memory. There are multiple nodes in the system, and each node has multiple pairs of CPU and GPU. CPUs and GPUs are hierarchically connected to out-node switches, and the out-node switches are connected to multiple remote memory groups. Remote memory groups collectively work as a shared memory pool for all CPUs and GPUs. Let's assume that there are 16 nodes with 16 pairs of CPU and GPU in each node. In total, there are 256 CPUs and 256 GPUs. Additionally, we assume that there are four out-node switches and eight remote memory groups.

\begin{figure}[t!]
    \centering
    \includegraphics[width=\linewidth]{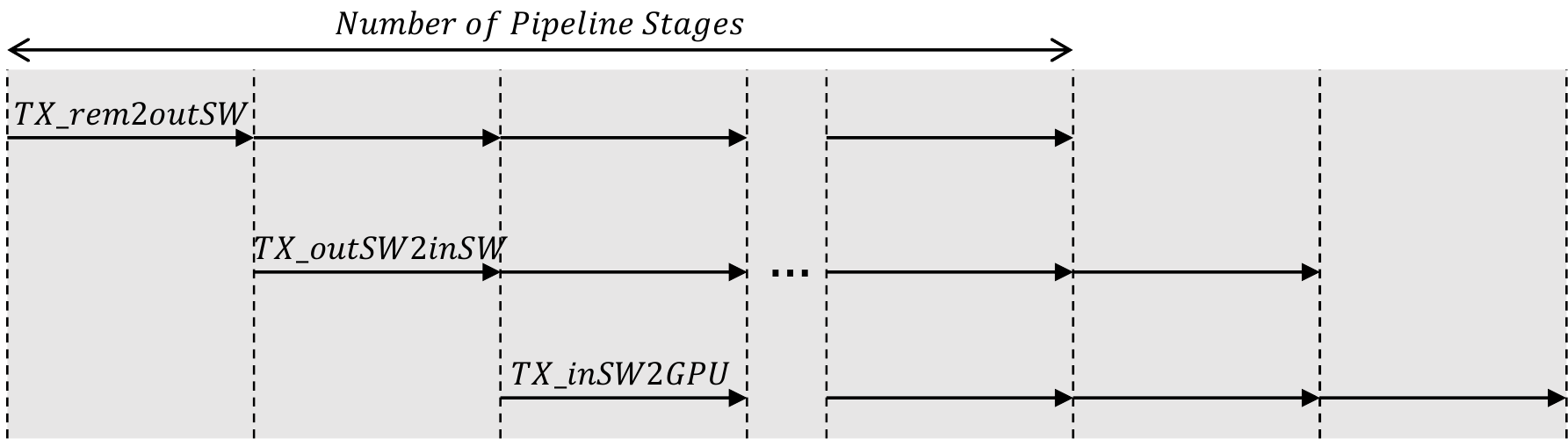}
    \caption{Pipelined data transfer.}
    \label{fig:pipelined-data-transfer}
    \vspace{-1em}
\end{figure}

If each GPU wants to load a tensor of size \texttt{W} from the remote memory pool, 256\texttt{W} should be loaded from the remote memory pool. As there are eight remote memory modules, each remote memory module will have 32\texttt{W}. As there are four out-node switches and each remote memory group is connected to all out-node switches, each link has to transfer 8\texttt{W}. The data to transfer on the link between an out-node switch and a node is 4\texttt{W} as each node requires 16\texttt{W} (the number of GPUs in a node) and four out-node switches will transfer the same amount of data. Once the loads on links are determined, they are transferred in a pipelined manner with the chunk size unit. The chunk size is the basic transfer unit of the network. \autoref{fig:pipelined-data-transfer} demonstrates how tensors are transferred in a pipelined manner. The meaning of notations is described as the following equations. The total data transfer time is the sum of the critical path, and the length of a stage is determined by the max of data transfer time (arrows) in the stage.

\vspace{-1.2em}
\begin{gather*}
\scalebox{.7}{$
    \begin{aligned}[b]
    &(Number\ of\ Pipeline\ Stages)\\
    &= ((Tensor\ Size)\times(Number\ of\ GPUs))\\
    &/ (Number\ of\ Remote\ Memory\ Groups)\\
    &/ (Number\ of\ Out{\text -}node\ Switches)\\
    &/ (Chunk\ Size)
    \end{aligned}$}
\end{gather*}
\vspace{-1em}
\begin{gather*}
\scalebox{.7}{$
    \begin{aligned}[b]
    &(TX\_rem2outSW)\\
    &= (Chunk\ Size) / (Mem{\text -}side\ Out{\text -}node\ Fabric\ BW)
    \end{aligned}$}
\end{gather*}
\vspace{-1em}
\begin{gather*}
\scalebox{.7}{$
    \begin{aligned}[b]
    &(TX\_outSW2inSW)\\
    &= ((Number\ of\ Remote\ Memory\ Groups)\times(Chunk\ Size))\\
    &/ ((Number\ of\ Nodes)\times(GPU{\text -}side\ Out{\text -}node\ Fabric\ BW))
    \end{aligned}$}
\end{gather*}
\vspace{-1em}
\begin{gather*}
\scalebox{.7}{$
    \begin{aligned}[b]
    &(TX\_inSW2GPU)\\
    &= ((Num\ of\ Rem\ Mem\ Groups)\times(Num\ of\ Out{\text -}node\ SW)\times(Chunk\ Size))\\
    &/ ((Number\ of\ GPUs)\times(In{\text -}node\ Fabric\ BW))
    \end{aligned}$}
\end{gather*}
\vspace{-1.2em}

\begin{figure}[t]
    \centering
    \includegraphics[width=\linewidth]{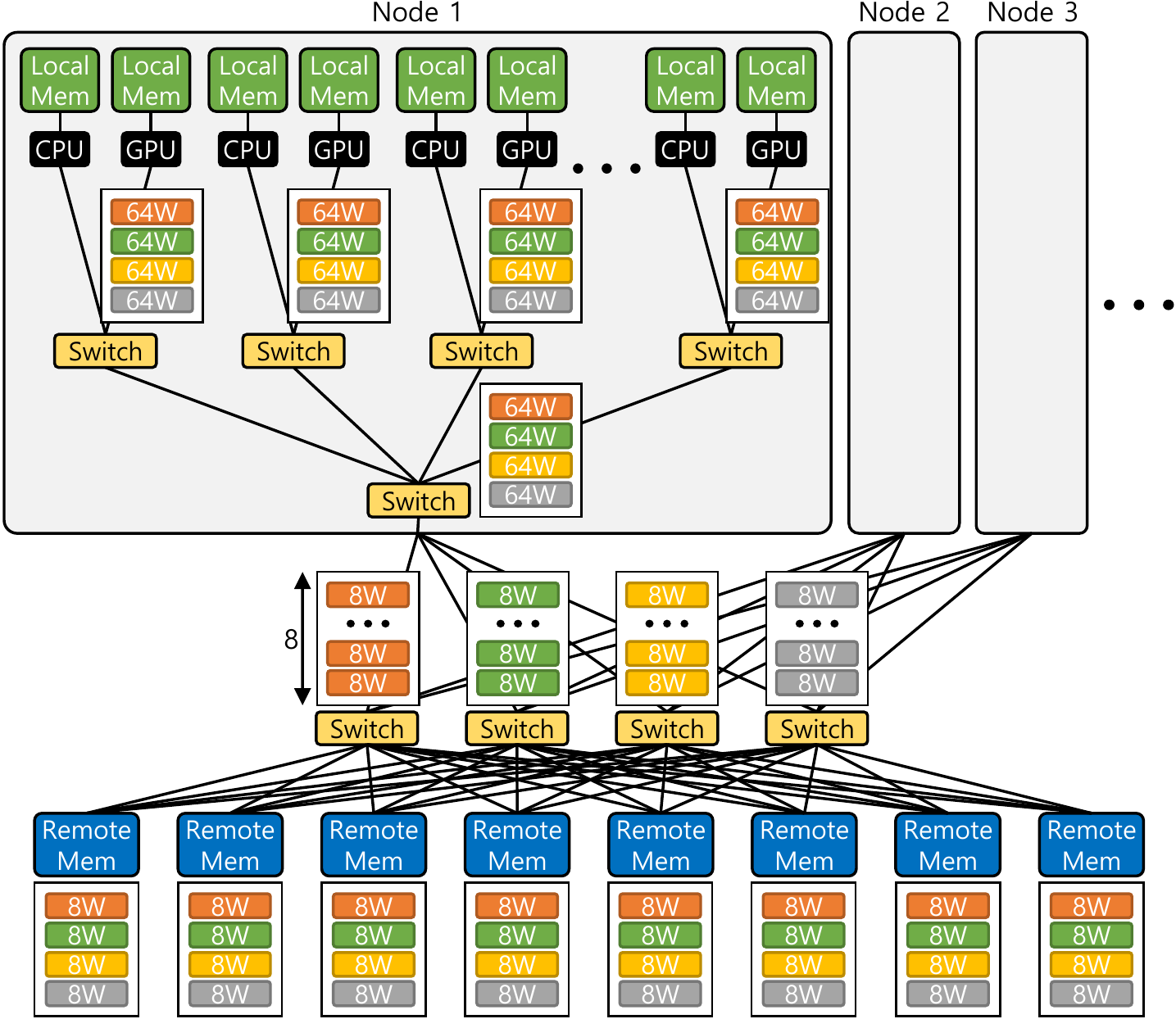}
    \caption{In-switch collective communication illustration.}
    \label{fig:in-switch-collective-communication-illustration}
    \vspace{-1em}
\end{figure}

\insertFigureWide{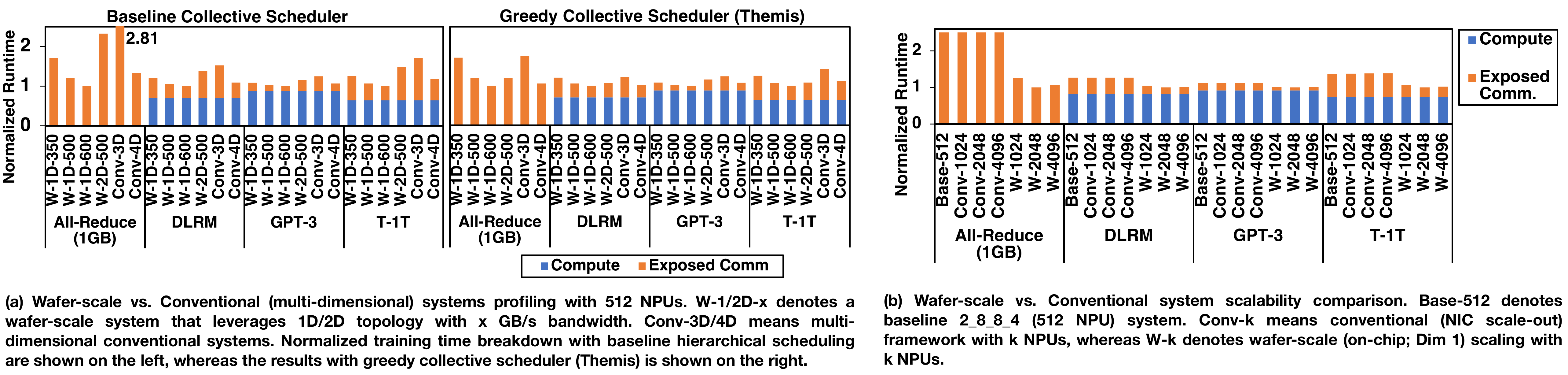}{
(a) Conventional (multi-dimensional) vs. wafer-scale training time breakdown, with and without greedy collective scheduling (Themis) policy. (b) Conventional (scale-out) vs. wafer-scale (on-chip) scalability analysis. Exposed Comm refers to the communication time that is not hidden behind compute time.
}{1}{-6mm}{-5mm}

\circled{3} \textbf{In-switch Collective Communication.} We support in-switch collective communication with an analytical model. With in-switch collective communication, parameters are gathered while being loaded (All-Gather), and sharded while being stored (Reduce-Scatter). The analytical model for in-switch collective communication is similar to the analytical model for remote memory access. However, the only difference is the data size to transfer for each link as parameters are gathered or scattered. Let's take the same example used for the remote memory model. \autoref{fig:in-switch-collective-communication-illustration} illustrates how in-switch collective communication works. In this figure, we assume that each GPU loads a tensor size of \texttt{W}. As there are 256 GPUs, the total size of tensors to load is 256\texttt{W}. The tensors are sharded into eight remote memory groups, and each remote memory group has 32\texttt{W}. As each remote memory group is connected to four out-node switches, each link transfers 8\texttt{W}. Each out-node switch will have  64\texttt{W} in total because eight remote memory groups transfer 8\texttt{W} for all out-node switches. While receiving the weights, they are gathered. After that, the out-node switches are forwarding 64\texttt{W} to each node. As a result, each in-node switch receives 256\texttt{W}, which is the reconstructed weight. In-node switches are responsible for broadcasting the gathered weights to GPUs. Parameters are transferred in a pipelined manner as shown in the remote memory model. In-switch collective communication changes the equations as below.

\begin{gather*}
\scalebox{.7}{$
    \begin{aligned}[b]
    &(TX\_rem2outSW)\\
    &= (Chunk\ Size) / (Mem{\text -}side\ Out{\text -}node\ Fabric\ BW)
    \end{aligned}$}
\end{gather*}
\vspace{-1.5em}
\begin{gather*}
\scalebox{.7}{$
    \begin{aligned}[b]
    &(TX\_outSW2inSW)\\
    &= ((Number\ of\ Remote\ Memory\ Groups)\times(Chunk\ Size))\\
    &/ (GPU{\text -}side\ Out{\text -}node\ Fabric\ BW)
    \end{aligned}$}
\end{gather*}
\vspace{-1.5em}
\begin{gather*}
\scalebox{.7}{$
    \begin{aligned}[b]
    &(TX\_inSW2GPU)\\
    &= ((Num\ of\ Rem\ Mem\ Groups)\times(Num\ of\ Out{\text -}node\ SW)\times(Chunk\ Size))\\
    &/ (In{\text -}node\ Fabric\ BW)
    \end{aligned}$}
\end{gather*}

%% file: table/TopologyAwareCollective.tex
\begin{table}[t]
    \footnotesize{
\caption{Network building blocks and its corresponding topology-aware collective communication algorithm~\cite{nccl}.}
\vspace{-3mm}
\begin{center}
\resizebox{\columnwidth}{!}{%
    \begin{tabular}{|c|c|}
    \hline
    \textbf{Network Building Block} & \textbf{Topology-aware Collective} \\ \hline
    \ring              & \ringAlg~\cite{ringCollective}               \\ \hline
    \fc   & \directAlg~\cite{rabenseifner}             \\ \hline
    \switch            & \hdAlg~\cite{rabenseifner}    \\ \hline
    \end{tabular}
}
\vspace{-7mm}
\label{table:TopologyAwareCollective}
\end{center}
}
\end{table}

%% file: content/case_studies.tex
\section{Case Studies}
In this section, we run comprehensive case studies showcasing the extended capabilities of ASTRA-sim and provide meaningful insights regarding the design space. For all our experiments, we assumed NPU compute power of 234 TFLOPS observed from the measurements of an A100 GPU~\cite{dgxa100}.

\subsection{Conventional System vs Wafer-scale System}

Wafer-scale systems feature a number of NPUs (on a wafer) connected using low-dimensional but high-BW on-chip (on-wafer) interconnection networks~\cite{packageLessProcessing,cerebraswhitepaper, dojo}. Meanwhile, conventional systems~\cite{dgxa100,cloudtpuarch} have multi-dimensional hierarchical topologies with various networking techniques including on-chip, scale-up, and scale-out (NIC). We compare the two distinct approaches by abstracting these systems. Target experimental topologies with 512 NPUs are summarized in \autoref{table:TopologySetup}. For wafer-scale proxy, we create three 1D topologies with 300, 500, and 600 GB/s on-wafer BW (W-1D), and a 2D topology with 250\_250 GB/s BW (W-2D) to model futuristic wafer systems~\cite{packageLessProcessing,waferScaleGPU}. For conventional systems, we created 3D and 4D topologies (Conv-3D and Conv-4D) using on-chip, scale-up, and scale-out interconnections borrowed from~\cite{dgxa100,nvidiadgx,themis}. Target distributed training workloads and their characteristics are also summarized in \autoref{table:WorkloadSetup}.
\input{table/TopologySetup.tex}

\input{table/WorkloadSetup.tex}

\subsubsection{\textbf{Impact of Scheduling}}
Normalized runtimes of a single 1GB \allreduce as well as real workloads are shown in \autoref{fig:ResultWaferScale}. When a topology is multi-dimensional, complex behaviors like pipelining bubbles or unbalanced network BW result in low BW resource utilization and sub-optimal performance~\cite{themis}. Having only one dimension, \textit{\textbf{W-1D yielded the overall best performance}}. However, if you specifically compare W-1D-350 and Conv-4D (600GBps/NPU), \textit{\textbf{Conv-4D is driving more BW/NPU, showing better performance}} despite being multi-dimensional.
Next, we study the impact of scheduling. Themis is a greedy scheduling policy for collectives that aims to balance the load across multiple dimensions to achieve near-optimal BW utilization~\cite{themis}. W-1D topologies, already being only 1D, show no gain from smart scheduling as shown in \autoref{fig:ResultWaferScale}(a). However, W-2D, Conv-3D, and Conv-4D, being \textit{\textbf{multi-dimensional, heavily benefit from Themis scheduler}}. It is worth noting that for single \allreduce and \dlrm, \textit{\textbf{conventional systems with Themis scheduler shows identical results compared to its corresponding wafer-scale systems with equivalent BW/NPU}}. Considering the complexity and cost to build a system on a single wafer, such results glimpse the possible advantage of using the conventional hierarchical approach in performance-per-cost aspects. Meanwhile, for \gpt and \tlarge, wafer-scale systems still maintained better training time. For hybrid parallelism on conventional systems, MP and DP spans over some (and not every) dimensions and utilize only those BW, whereas for wafer-scale every communication runs on full on-wafer BW. This emphasizes the importance of \textit{\textbf{appropriate parallelization strategies and the need to co-design them with underlying topologies}} for conventional hierarchical systems.

\subsubsection{\textbf{Impact of Scaling using Wafer-scale Systems}}
\input{table/DimMessageSize.tex}
Traditional systems scale the infrastructure by scale-out approach, i.e., attach more nodes to the last-dim NICs. On the contrary, wafer-scale technologies let the framework scale up the system, i.e., increasing the number of NPUs on-chip (Dim 1) while maintaining the number of scale-out nodes equally. Measuring the impact, we take the Conv-4D topology from \autoref{table:TopologySetup}, set the on-chip (i.e., Dim 1) BW to 1,000GB/s to model wafer-scale systems~\cite{packageLessProcessing,waferScaleGPU}, and set it as a baseline. Then, we scale the platform up to 4K nodes and measured the 1GB \allreduce time. The results are shown in \autoref{table:DimMessageSize}. \textit{\textbf{Conventional scale-out increases the Dim 4 (NIC) message size, but the impact was marginal}}, thereby showing identical collective time. \textit{\textbf{Scaling over the wafer, however, significantly increased on-wafer (Dim 1) communication size while dramatically cutting down other dimensions' load}}. As far as the system has enough on-wafer BW, collective time decreases due to such an effect, showing an up to 2.51$\times$ speedup over the corresponding scale-out mechanism. \textit{\textbf{Once the on-wafer dimension becomes the bottleneck, the collective time starts to bounce and increase again}} as can be seen from the 16\_8\_8\_4 system. End-to-end training time breakdown of \gpt and \tlarge is also shown in \autoref{fig:ResultWaferScale}(b), showing the equivalent trend in the end-to-end regime.

\subsection{Comparing Disaggregated Memory Systems}

In this case study, we compare the performance of two disaggregated memory systems: ZeRO-Infinity~\cite{rajbhandari2021zero} and the hierarchical memory pool (\texttt{HierMem}) presented in \autoref{sec:memory-models}. We compare the performance of the disaggregated memory systems because the latest model sizes already exceed the memory capacity of GPUs available in the market. ZeRO-Infinity is chosen as a baseline system as it is proposed as an effective solution to overcome the limited memory capacity. ZeRO-Infinity is a nascent form of memory disaggregation where each GPU can utilize CPU memory and NVMe in addition to its local HBM memory. \autoref{fig:ZeRO_Infinity_system_design} presents the system architecture of ZeRO-Infinity. While ZeRO-Infinity has an advantage in terms of its availability in commodity servers, it does not allow having an arbitrary number of remote memory groups. In other words, it cannot enjoy the major benefit of memory disaggregation, which is cost reduction by eliminating memory underutilization. On the other hand, \texttt{HierMem} can have an arbitrary number of remote memory groups. System parameters for the baseline \texttt{HierMem} configuration are presented in \autoref{tab:case-study-disagg-mem-system-configurations}. The values for the baseline configuration are determined based on the latest GPU performance and network bandwidth of commodity servers.

\insertFigure{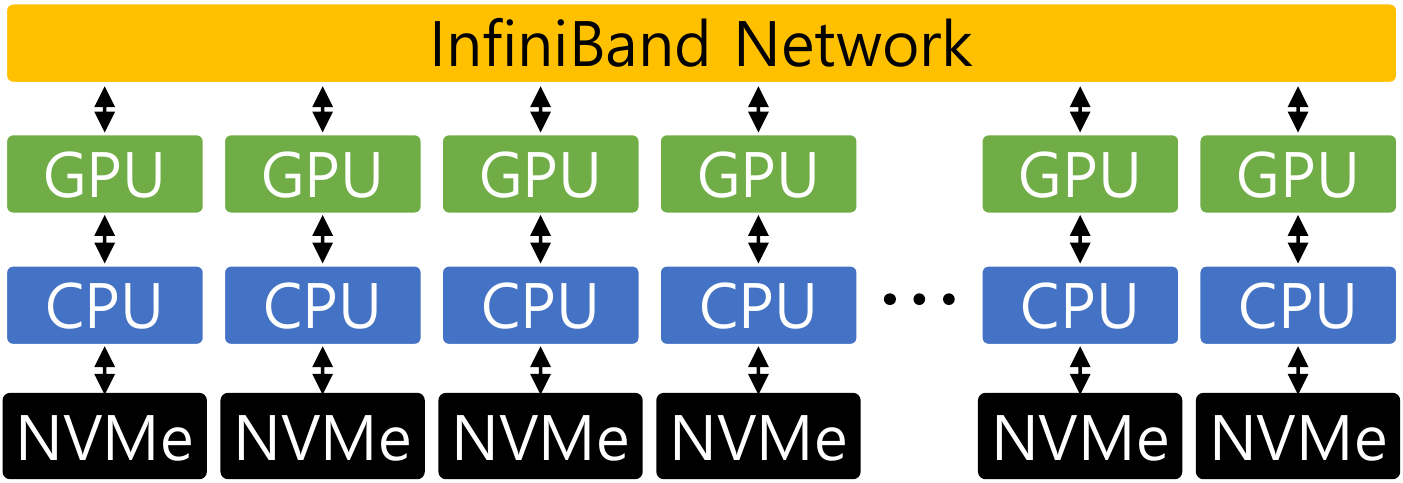}{
ZeRO-Infinity system architecture.
}{0.75}{-2mm}{-1mm}

\begin{table}[t!]
    \center
    \caption{Disaggregated memory system configurations}
    \scalebox{0.8}{
    \begin{tabular}{|l|r|r|r|}
    \hline
                                               & \multicolumn{1}{c|}{\textbf{ZeRO-Infinity}} & \multicolumn{1}{c|}{\textbf{\begin{tabular}[c]{@{}c@{}}HierMem\\ (Baseline)\end{tabular}}} & \multicolumn{1}{c|}{\textbf{\begin{tabular}[c]{@{}c@{}}HierMem\\ (Opt)\end{tabular}}} \\ \hline
    \textbf{GPU Peak Perf (TFLOPS)}            & 2048                                        & 2048                                                                                       & 2048                                                                                  \\ \hline
    \textbf{GPU Local HBM BW (GB/sec)}         & 4096                                        & 4096                                                                                       & 4096                                                                                  \\ \hline
    \textbf{In-node Pooled Fabric BW (GB/sec)} & -                                           & 256                                                                                        & \textbf{512}                                                                          \\ \hline
    \textbf{Num of Out-node Switches}          & -                                           & 16                                                                                         & 16                                                                                    \\ \hline
    \textbf{Num of Remote Memory Groups}       & 256                                         & 256                                                                                        & 256                                                                                   \\ \hline
    \textbf{Remote Mem Group BW (GB/sec)}      & 100                                         & 100                                                                                        & \textbf{500}                                                                          \\ \hline
    \end{tabular}}
    \label{tab:case-study-disagg-mem-system-configurations}
    \vspace{-1.3em}
\end{table}

To compare the performance of the systems, we run a training task for a mixture-of-experts (MoE) model with 1 trillion parameters~\cite{rajbhandari2022deepspeed}.
\autoref{fig:execution_time_breakdown_comparison} presents the execution time breakdown.
The execution time of a training task can be broken down into five components: compute time, exposed local memory access time, exposed remote memory access time, exposed communication time, and exposed idle time. The compute time is the total compute time to train a model, and other operations can be hidden behind each other. Non-hidden time of an operation is defined as exposed time. Overall, ZeRO-Infinity performs 0.1\% better than \texttt{HierMem}. Both memory systems present similar performance because they have almost equivalent resources. The small performance drop in \texttt{HierMem} originates from the additional data transfer stages with multi-level switches.

To find a better-performing configuration of \texttt{HierMem}, we explore the design space of \texttt{HierMem} while varying in-node pooled fabric bandwidth and the remote memory group bandwidth. We only sweep these parameters as the exposed communication turns out to be a bottleneck. In-node pooled fabric bandwidth is varied between 256GB/s and 2048GB/s with the unit of 256GB/s, and remote memory group bandwidth is varied between 100GB/s and 500GB/s with the unit of 100GB/s. The found best performance with the least resource provision is shown as \texttt{HierMem(opt)} in \autoref{tab:case-study-disagg-mem-system-configurations} and \autoref{fig:execution_time_breakdown_comparison}. It performs 4.6 times better than the baseline configuration.

\insertFigure{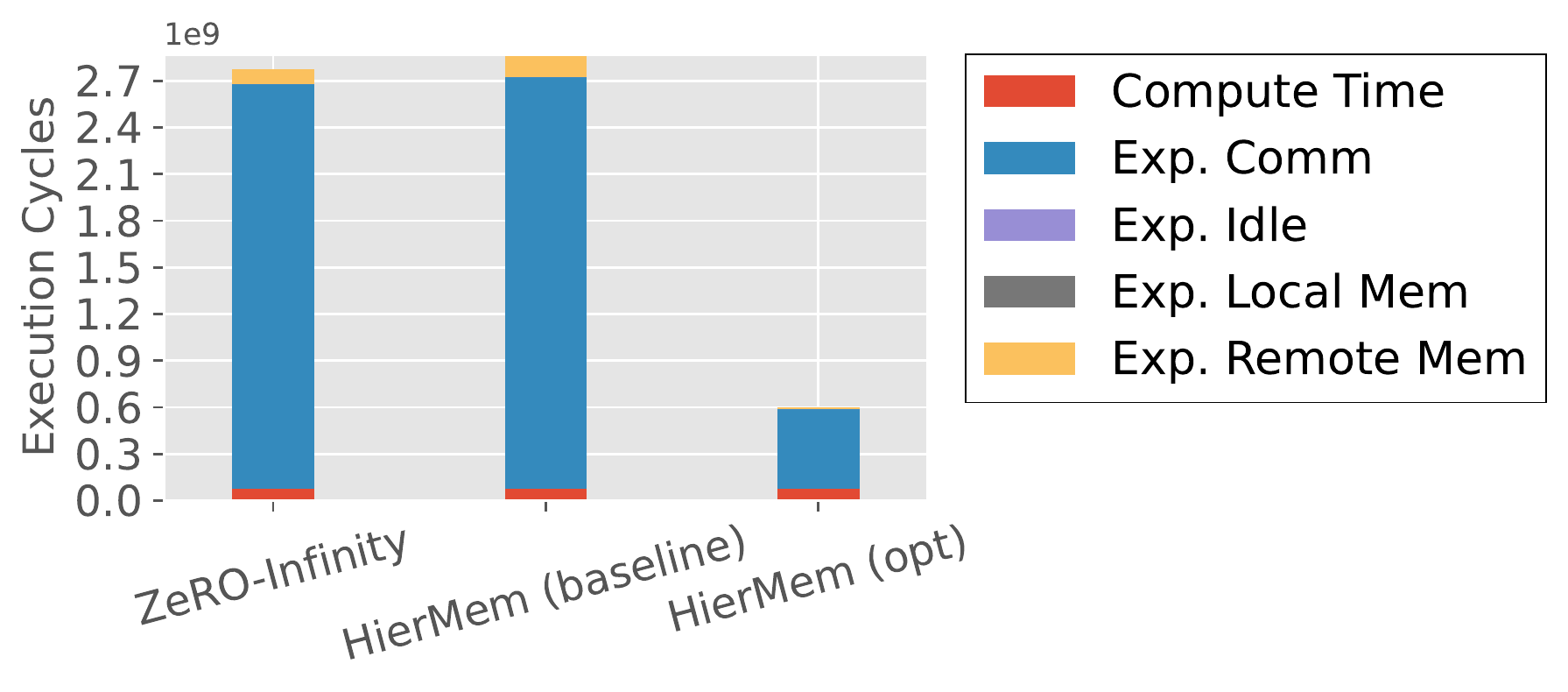}{
Runtime breakdown of disaggregated architectures.
}{0.94}{-3mm}{-4mm}

%% file: table/TopologySetup.tex
\begin{table}[t]
\small
\caption{Target wafer-scale and conventional (multi-dimensional) topologies. Conventional system parameters are borrowed from \cite{dgxa100,nvidiadgx} and wafer-scale params are borrowed from \cite{packageLessProcessing,waferScaleGPU}.}
\vspace{-3mm}
\begin{center}
\resizebox{0.9\columnwidth}{!}{%
    \begin{tabular}{|c|c|c|c|}
    \hline
    \textbf{Topology} & \textbf{Shape} & \textbf{NPU Size} & \textbf{BW (GB/s)} \\ \hline
    W-1D & Switch & 512 & 350, 500, 600 \\ \hline
    W-2D & Switch\_Switch & 32$\times$16 & 250\_250 \\ \hline
    Conv-3D & Ring\_FC\_Switch & 16$\times$8$\times$4 & 200\_100\_50\\ \hline
    Conv-4D & Ring\_FC\_Ring\_Switch & 2$\times$8$\times$8$\times$4 & 250\_200\_100\_50\\ \hline
    \end{tabular}
}
\vspace{-4mm}
\label{table:TopologySetup}
\end{center}
\end{table}

%% file: table/WorkloadSetup.tex
\begin{table}[t]
\small
\caption{Target training workloads and their characteristics.}
\vspace{-3mm}
\begin{center}
\resizebox{0.9\columnwidth}{!}{%
    \begin{tabular}{|c|c|c|c|}
    \hline
    \textbf{Workload} & \textbf{\#Params} & \textbf{MP Size} & \textbf{DP Size} \\ \hline
    DLRM    & 57M (MLP layers) & 1,024              & 1,024                \\ \hline
    GPT-3   & 175B                 & 16               & 64               \\ \hline 
    Transformer-1T    & 1T               & 128              & 8                \\ \hline
    \end{tabular}
}
\vspace{-7mm}
\label{table:WorkloadSetup}
\end{center}
\end{table}

%% file: table/DimMessageSize.tex
\begin{table}[t]
\small
\caption{Message size (MB) per each dimension and collective time when running an 1GB \allgather collective.}
\vspace{-3mm}
\begin{center}
\resizebox{\columnwidth}{!}{%
\begin{tabular}{|c|r|r|r|r|r|r|}
\hline
\textbf{\begin{tabular}[c]{@{}c@{}}System\\ Size\end{tabular}} & \textbf{NPUs} & \textbf{Dim 1} & \textbf{Dim 2} & \textbf{Dim 3} & \textbf{Dim 4} & \textbf{\begin{tabular}[c]{@{}r@{}}Collective\\ Time (µs)\end{tabular}} \\ \hline
2\_8\_8\_4      & 512           & 1024           & 896            & 112            & 12             & 4392.85                                                                 \\ \hline
2\_8\_8\_\textbf{8}      & 1,024         & 1024           & 896            & 112            & 14             & 4392.85                                                                 \\ \hline
2\_8\_8\_\textbf{16}     & 2,048         & 1024           & 896            & 112            & 15             & 4392.85    \\ \hline
2\_8\_8\_\textbf{32}     & 4,096         & 1024           & 896            & 112            & 15.5             & 4392.85    \\ \hline
\textbf{4}\_8\_8\_4      & 1,024         & 1536           & 448            & 56             & 6              & 2212.60                                                                 \\ \hline
\textbf{8}\_8\_8\_4      & 2,048         & 1792           & 224            & 28             & 3              & 1753.48                                                                 \\ \hline
\textbf{16}\_8\_8\_4      & 4,096         & 1920           & 112            & 14             & 1.5              & 1879.17                                                                 \\ \hline
\end{tabular}

}
\vspace{-7mm}
\label{table:DimMessageSize}
\end{center}
\end{table}

%% file: content/related_work.tex
\section{Related Work}
Several simulators exist in our community for modeling distributed systems running general-purpose workloads~\cite{mohammad2017dist, sst, zsim}, with the classic trade-off between simulation accuracy, simulation speed and engineering effort.
Moreover, several models/simulators have been proposed to optimize communication performance in HPC platforms, such as LogGOPSim~\cite{Hoefler2010loggopsim} and SMPI~\cite{Degomme2017smpi}.
This work builds upon the observation of recent works~\cite{astrasim, Robinson2022DTS, deepflow, themis} that the compute-memory-communication characteristics of distributed training is possible to abstract and capture via a mix of analytical and simulation methods, without requiring a general-purpose simulator.
This is the first simulator, to the best of our knowledge, to enable running arbitrary DNN training execution traces over next-generation platforms with multi-dimensional (scale-up + scale-out) topologies and disaggregated memory systems.

%% file: content/conclusion.tex
\section{Conclusion}
In this paper, we motivate the need to swiftly model and profile state-of-the-art and emerging training platforms running large DL models. We enhance the capabilities of \astrasim to enable capturing arbitrary parallelization strategies and training loops, supporting multi-dimensional network topologies, and representing complex memory systems. Using the framework, we run a comprehensive end-to-end, full-stack co-design space exploration of distributed training. With the ability to quickly navigate the complex design space of distributed training, this can give meaningful first-order insights to system designers and assist them in building futuristic training platforms at scale.

%% file: include/acknowledgement.tex
\section*{Acknowledgment}
This work was supported by awards from Intel and Meta. The tool is being maintained via support from Semiconductor Research Corporation.